\documentclass[10pt,letterpaper,twocolumn]{article} 

\usepackage{ol2}
\usepackage[draft]{hyperref}
\usepackage{amsmath}

\begin{document}

\twocolumn[ 

\title{Storage of orbital angular momenta of light via coherent population oscillation}


\author{A.~ J.~ F.~ de Almeida${^1}$, S.~Barreiro$^{1,\dagger}$, W.~S.~Martins$^{1}$, R.~A.~de Oliveira$^{1,\ddagger}$
D.~Felinto$^{1}$, L. ~Pruvost${^2}$, and J.~W.~R.~ Tabosa${^{1,*}}$}

\address{
$^1$Departamento de F\'{\i}sica, Universidade Federal de Pernambuco, 50670-901 Recife, PE, Brazil \\
$^2$Laboratoire Aim\'{e}  Cotton, CNRS, Universit\'{e} Paris Sud, ENS Cachan, 91405 Orsay, France \\
$^*$Corresponding author: tabosa@df.ufpe.br
}

\begin{abstract}
We report on the storage of Orbital Angular Momentum (OAM) of light via the phenomenon
of Coherent Population Oscillation (CPO) in cold cesium atoms. The experiment is
performed using a delayed four wave mixing configuration where the transverse
optical information of a probe field carrying OAM associated with its azimuthal
phase dependence is stored in the CPO of Zeeman sublevels of the hyperfine 
transition $F=3\rightarrow F^{\prime}=2$ of cesium $D_{2}$ line. We also demonstrate
experimentally the simultaneous storage and retrieval of different OAM states
propagating along different directions in space, leading to algebric operations with OAM and therefore opening the possibility
of multiplexing OAM states. 
\end{abstract}

\ocis{270.1670, 190.4223, 020.1670}

 ] 

\noindent An optical memory, i.e., the possibility of 
storing information carried by a light beam into an atomic medium
for later on demand retrieval, is in the heart of any classical or quantum protocol for processing and manipulating
light information  \cite{Lukin03, Lvovsky09}.
Many optical memories are based on the  reversible transfer of
coherence between light and matter achieved through different physical phenomena
such as  Electromagnetically Induced Transparency (EIT) \cite{Hau01, Lukin01},
Gradient Echo Memories (GEM) \cite{Hosseini09, Buchler12}, and  Atomic Frequency
Comb (AFC) \cite{Afzelius09}, which use long-lived ground-state coherence to store
the optical information.
  
Recently, a new type of optical memory based on the CPO phenomenon \cite {Boyd81, Berman88,
Goldfarb12} has been suggested \cite{Wilson-Gordon10} and then demonstrated
\cite{Allan14, Goldfarb14}, using intense light fields. The probe absorption spectrum of a driven-two-level atomic system presents CPO resonances around the zero beating frequency, which are associated with the dynamics of atomic level populations. Not based on ground-state coherence, as an EIT-based memory, the CPO memory uses the long relaxation time of the ground state of an open two-level system to store information carried by a light field.
In the demonstrated CPO memory,  the population oscillation occurs into two
open two-level systems coupled by spontaneous emission, where each open two-level
system is excited simultaneously by the different field polarization components.  
Unlike the EIT memory, the insensibility to magnetic field inhomogeneities 
makes the CPO-based memory ideal to store spatial information and finds
new insights for storing and manipulating  orbital angular momentum (OAM) of light.

Light modes with topological charge, Laguerre-Gauss (LG) modes for
example, have a spatial dependence through an azimuthal phase given by
$e^{i\ell\phi}$ where $\ell $ is an integer, usually called the topological charge. They carry a quantized OAM $\ell
\hbar$  per photon \cite{Allen92} and are usable for information processing based on the
multidimensional state space spanned by these modes
\cite{Bechmann-Pasquinucci00}. Optical storage of OAM has  been demonstrated previously, both in
the classical \cite{Pugatch07, Moretti09} and in the quantum single-photon
\cite{Laurat13, Guo13, Guo14} regimes employing the EIT based memories.
Recently the storage and retrieval of photonic qubits carrying
OAM was also demonstrated \cite{Laurat14}.

In this Letter we demonstrate the storage and retrieval of OAM of light using
the CPO mechanism in an ensemble of cold cesium atoms. This result is the first demonstration of storage and retrieval of the transverse spatial-phase structure of an optical field using this new type of memory. Moreover, we demonstrate the co-storage of two OAM states
carried by two optical beams and the retrieval of the
sum of the associated topological  charges.

The experiment has been performed in cold cesium atoms provided by a
magneto-optical trap (MOT) in which two incident
beams W, W' write the information (storage) and another one R reads it (retrieval)
according to the backward four-wave mixing (FWM) geometry shown in Fig. 1 (a). The retrieved information is delivered by the diffracted beam denoted
by D.
The optical memory uses the degenerate two-level system associated with the cesium hyperfine transition
$6S_{1/2}(F=3)\leftrightarrow 6P_{3/2}(F^{\prime}=2)$, as shown in Fig. 1(b). 
To prepare the atoms in the lower hyperfine ground state $F=3$, we switch off
the MOT repumping beam $1$ $ms$ before switching off the MOT trapping beams. The atomic cloud optical depth is of the order of 4. All the laser beams involved in the experiment are provided by two amplified extended-cavity diode lasers, and are time and frequency controlled using acousto-optics
modulators (AOM) in double passage configuration.The MOT magnetic quadrupole field is also turned off during the
storing and reading procedures.

In a first experiment, we turn on simultaneously the writing W, W'
and the reading R beams. All the incident beams have approximately the same diameter
of the order of $1.2 $ $mm$ which is smaller than the trapped cloud size ($\approx 2 $ $mm$). 
The intensities of the beams W, W' and R are respectively equal to $10$ $mW/cm^2$, $15$ $mW/cm^2$ and $5 $ $mW/cm^2$. 
The writing beams W and W' make a small angle of $\theta \approx 2^o$ between them. The reading beam R counter-propagates W, so the
diffracted conjugate beam, D, counter-propagates W', as imposed by phase matching
condition. D is detected by a fast photodiode. A dc longitudinal magnetic field of $B\approx 3.7$ $G$ is applied along
the direction of propagation of the beams, which defines the quantization axis and
Zeeman shifts consecutive ground sublevels by $\approx 1.3$ $ MHz$. To get the spectra shown in Fig.1(c),
the W laser is locked to the $F=3\leftrightarrow F^{\prime}=2 $ transition, while W'
and R are frequency-scanned through the frequency detuning $\delta$ defined in
Fig. 1(b). To satisfy the CPO configuration, W and W' are set linearly polarized, with
orthogonal polarization, denoted by $ lin \perp lin$. The FWM spectrum (black line of Fig.
1(c)) exhibits a narrow CPO resonance at $\delta =0$ and two weaker and broader
ones located at $\delta \approx \pm 2.5$ $MHz$ corresponding to EIT resonances. In order to check
and compare, the FWM spectrum under the same experimental conditions, but with orthogonal circularly
polarized writing beams has also been recorded. As shown by the red curve in Fig 1(c), which is ten-fold magnified, just one single peak is present around the two-photon Raman resonance (EIT) and no CPO resonance at $\delta=0$ is observed in such case. The FWM reflectivity for the CPO and EIT peaks are of the order of $0.3\%$ and $0.01\%$, respectively.

\begin{figure}[!tbp]
  \centerline{\includegraphics[width=11 cm, angle=0]{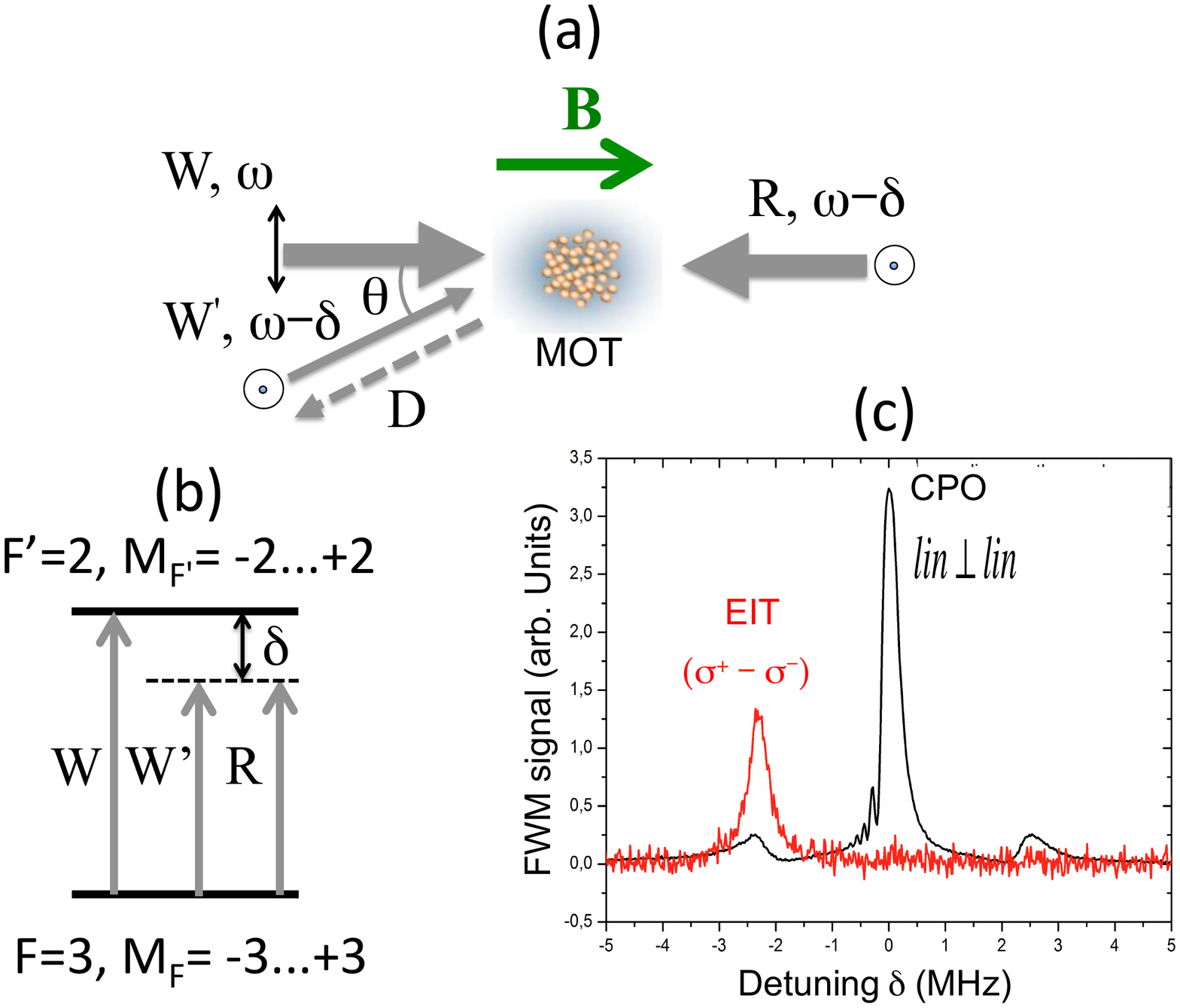}}
  \vspace{-0.5cm}
  \caption{(Color online) (a): Experimental scheme for the observation of CPO/EIT FWM in cold cesium
atoms. (b) Degenerate hyperfine two-level transition of cesium $D_{2}$ line involved in the
interaction with W,  W' and R beams. (c) CPO/EIT FWM spectral responses recorded on the diffracted D beam. Black curve:
FWM spectrum for orthogonal linear polarization of the writing beams; the peak at
$\delta=0$ is associated with the CPO resonance, while the lateral ones are associated with EIT resonances. Red
curve: FWM spectrum (x10) for orthogonal circular polarizations of the writing beams,
exhibiting only one peak associated with EIT resonance. }
  \label{fig:Fig1}
\end{figure}

In a second experiment, we have checked the robustness of the CPO memory against magnetic field inhomogeneities by performing a light storage measurement using the time sequence of Fig. 2 (a) where the R beam is switched on with a delay $t_s$ after the
switching off of W, W' beams and in presence of a magnetic field gradient. For that, we keep a reduced MOT quadrupole magnetic field in
order to create a magnetic gradient of $\approx 0.2$ $ G/cm$ during the storage/retrieval sequence. To distinguish CPO from EIT we use the polarization configurations and the dc longitudinal magnetic field at a detuning $\delta=0$. For the CPO memory we use the $lin \perp lin$ configuration in the presence of the applied longitudinal dc magnetic field, while for the EIT memory the dc magnetic field is turned off and the polarizations of W, W' beams are made orthogonal circularly polarized. The retrieved pulse is then observed for different storage times $t_s$. The retrieved pulse amplitude as a function of the storage time $t_s$  is shown in Fig. 2 (b) for the CPO and EIT memories and gives a storage decay time, respectively, equal to $9.2$ $ \mu s$ and $5.1$ $ \mu s$. In the inset in Fig.2 (b) we show a typical retrieved pulse for the CPO memory. The estimated storage efficiency (at $t_{s}\approx 1\mu s$) for the CPO and EIT memories, evaluated by the ratio between the retrieved pulse peak intensity and the W' intensity, are $1\%$ and  $2\%$, respectively.

It is worth noticing that in the absence of the magnetic field gradient the two
memories present the same storage decay time. In any case, the storage decay time of the CPO
memory is, however, much shorter than the expected decay time of $120$ $ \mu s$ 
given by the transit time of the atoms through the $
\Lambda=\frac{\lambda} {2sin (\theta/2)}  \approx 25$ $ \mu m$ spatial structure created by W, W' beams.  We
attribute this discrepancy to the existence of spurious transversal magnetic field 
components which can induce transitions between the Zeeman sublevels, therefore decreasing the
effective ground state lifetime of the Zeeman sublevels.

\begin{figure}[!tbp]
  \centerline{\includegraphics[width=12cm, angle=0]{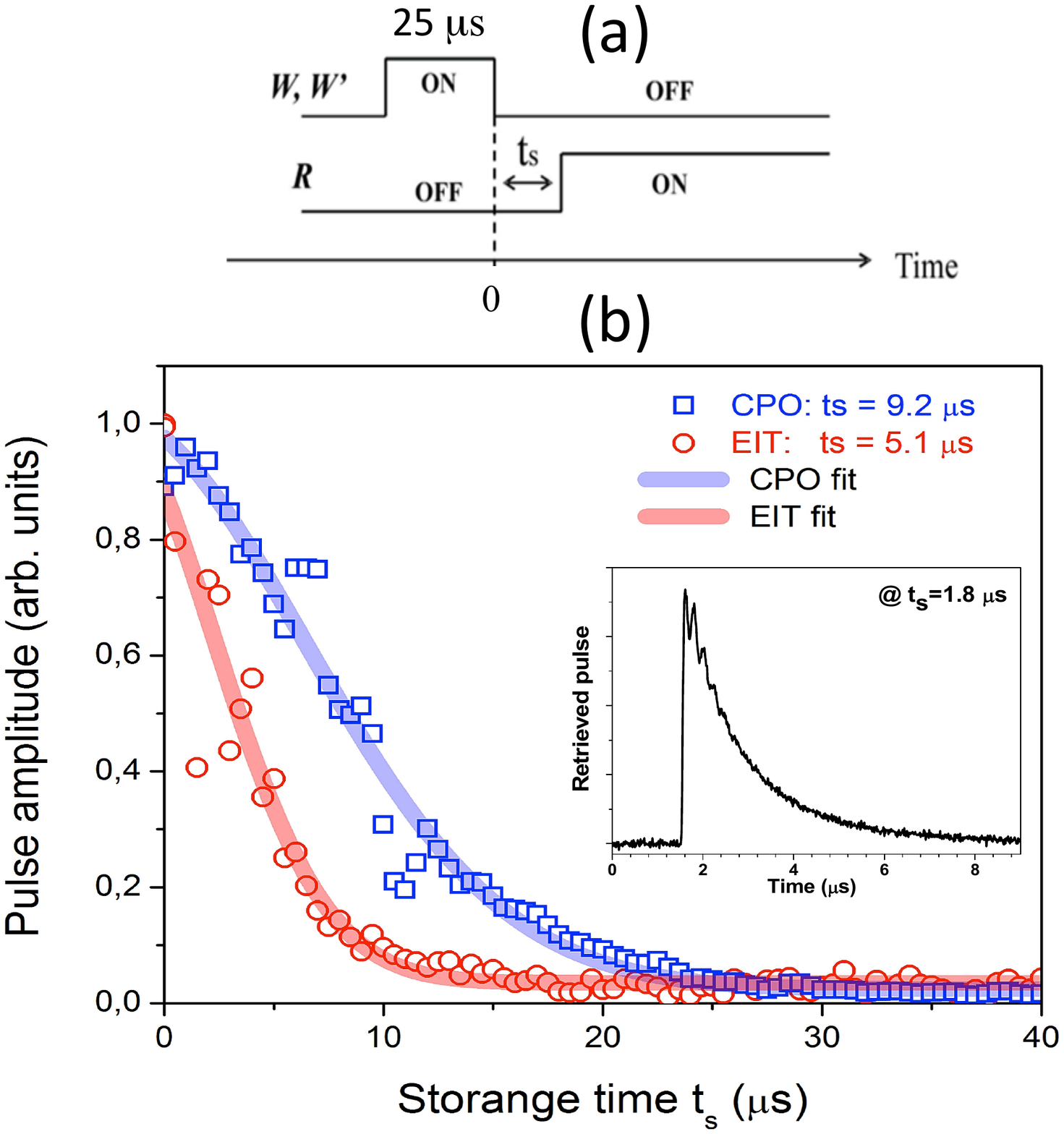}}
  \vspace{-0.7cm}
  \caption{(Color online) (a) Time sequence for the writing and reading beams.
(b) Decay of the retrieved pulse amplitude for CPO (blue squares) and EIT (red
circles) memories in the presence of a magnetic field gradient of $
\approx 0.2$ $ G/cm$. The solid lines correspond to the respective gaussian fittings. The inset shows a typical retrieved pulse for the CPO memory.}
  \label{fig:Fig2}
\end{figure}

In a third experiment, the beams W, W' have been phase-shaped into LG modes carrying OAM
with the respective topological charges $\ell_{W}$ an $\ell_{W'}$  in order to
analyze the CPO memory for OAM beams (Fig. 3 (a)).  So, W and W' are
frequency-locked on the $F=3\leftrightarrow F^{\prime}=2 $ transition ($\delta=0$) in a
$lin \perp lin$ polarization configuration. The shaping uses the holographic method where an
helical phase is imprinted on a Gaussian beam.  The hologram is provided by a Spatial
Light Modulators (SLM) as described in a previous paper \cite{Rafael14}
except  that we use now two identical SLMs. The diameter of the LG ring has been
adjusted to be slightly smaller than the MOT size. 
In order to check the topological charge of  the W and W' writing beams or to measure the
charge of the retrieved D beam, we used the tilted lens method described in
\cite{Singh13}. According to this method, the self-interference pattern of a LG beam
with topological charge $\ell$, observed after the astigmatic optics, exhibits
$|\ell|+1$  bright fringes turned by an angle whose sign is the sign of $\ell$. The
left column of Fig. 3 (b)  shows the self-interference pattern for  W'  with the
charge $\ell_{W'}= -2, -1, 0, +1, +2 $.  According to FWM phase-matching, for $\ell_{W}=  0 $,  the diffracted beam D acquires the topological charge
$\ell _{out}=\ell_{W'}$, that is illustrated by the right column of Fig. 3(b). 

In order to understand these patterns, we note that in the non saturating regime the
third-order nonlinear term of the optical polarization, responsible for the
generation of the retrieved field, propagating in opposite direction to the beam W',
has an amplitude given by: 
\vspace{-0.1cm}
\begin{equation}
{\cal{E}}_{D} \propto \chi^{(3)}{\cal{E}}_{W}{\cal{E}}_{W'}^{*}{\cal{E}}_{R},
\end{equation}
where, ${\cal{E}}_{i}$, for i=W, W', R are the amplitude of the incident fields, and $\chi^{(3)}$ the effective third-order susceptibility of the nonlinear medium. The phase
matching implies  both the wavevector conservation and the azimuthal phase
conservation. The writing beams W, W' are LG modes with an azimuthal phase given
by $\ell_{W}\phi$ and $\ell_{W'}\phi$, respectively and R is a gaussian mode. Thus,
for the considered FWM geometry, the azimuthal phase of ${\cal{E}}_{D}$ is
$(\ell_{W} -\ell_{W'})\phi$. However, due to wavevector conservation,  the D beam
counter-propagates W' and thus its topological charge defined relatively to its propagation direction is 
$\ell_{out}=\ell_{W'} -\ell_{W}$. Fig. 3 (b) illustrates this conservation law. 

In Fig. 3 (c) we have examined the off-axis OAM retrieval of the CPO memory, by
imposing $\ell_{W'}=0$ and $\ell_{W}= -2, -1, 0, +1, +2 $. We verify that
$\ell_{out}= -\ell_{W}$, as it is expected for small OAM values and small angles
between  W and W' \cite{Rafael14}. In both cases, on-axis and off-axis retrievals, we
have observed that the CPO storage decay time does not depend on the OAM values of the
writing beam. It is well known that the fork pattern associated with beams of W and W'
present a small region, near the center of the LG beam, where the fringes spacing changes as  $ \approx \lambda / (\ell +1)\theta$.
With $\ell =2$, the fringes spacing is reduced by a factor 3, reducing the transit time of
the atom in this structure by the same factor, so changing the previous expected
time to $40 \mu s$. This value still rests long compared to the observed CPO storage time. 
Larger OAM values could, however, affect the storage time.

\begin{figure}[!tbp]
  \centerline{\includegraphics[width=8 cm, angle=0]{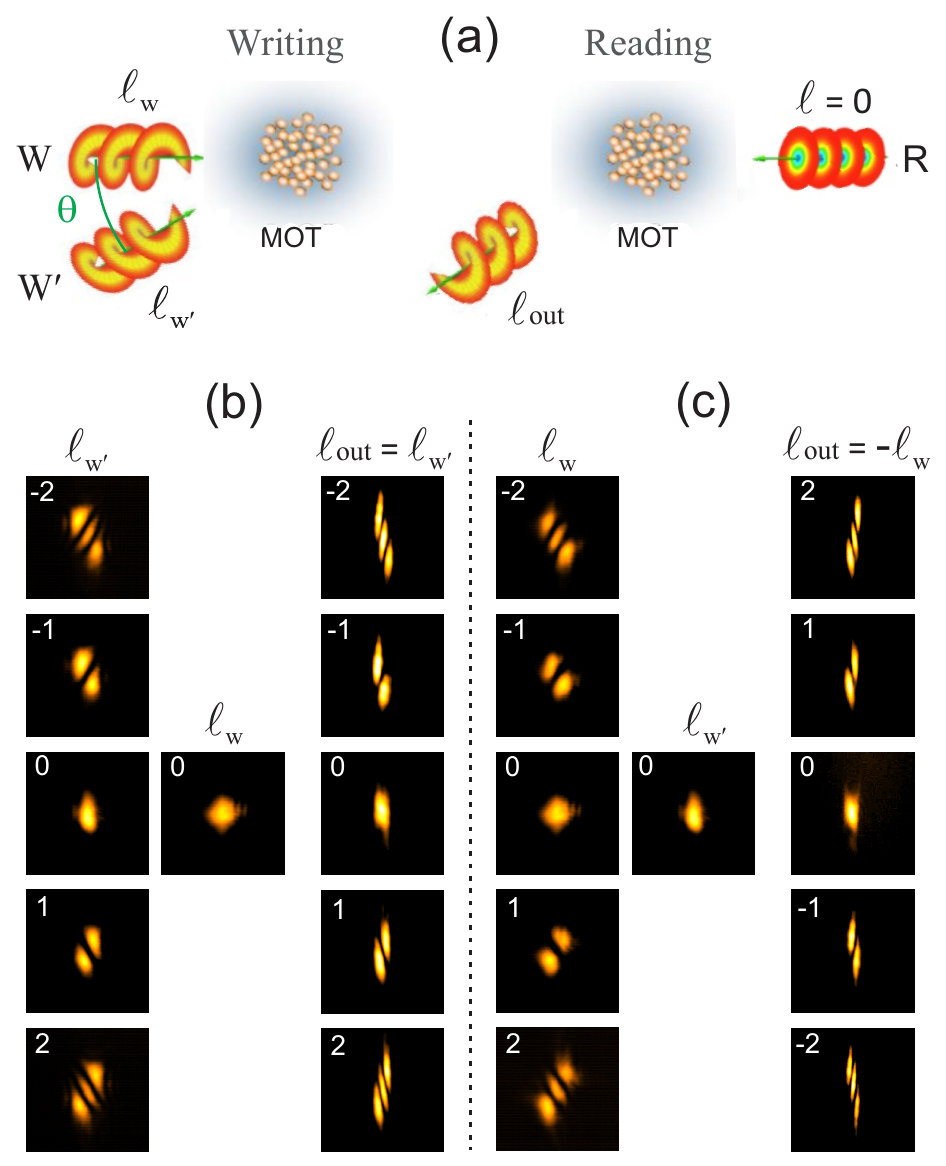}}
  \vspace{0.4cm}
  \caption{(Color online) (a): Experimental scheme for the writing and
reading process with OAM carrying writing beams. (b) and (c)
Topological charges of the input writing and retrieved beams recorded under CPO
conditions ($\delta=0$) after a storage time of $1 \mu s$ for $\ell_{W'}=0, \pm
1, \pm 2$ and  $\ell_{W}=0$ and for $\ell_{W}=0, \pm 1, \pm 2$, and $\ell_{W'}=0$, 
respectively. At the left up corner of each image we indicate the detected topological charge, either for the input and for the output beams.}
  \label{fig:Fig3}
\end{figure}

\begin{figure}[!tbp]
  \centerline{\includegraphics[width=7.5 cm, angle=0]{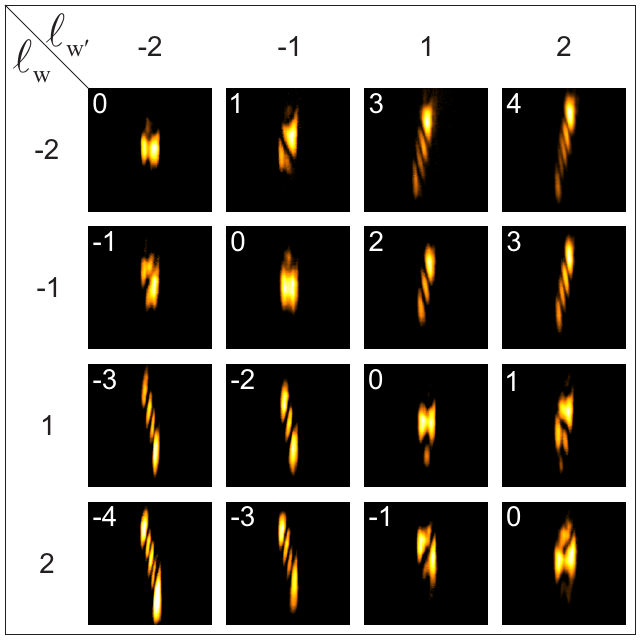}}
  \vspace{0.1cm}
  \caption{(Color online) Retrieved topological charge for different combinations of
non null values of $\ell_{W}, \ell_{W'}$. The value of the measured topological charge, $\ell_{out}$ is indicated at the left up corner of each frame.}
  \label{fig:Fig4}
\end{figure}

In another series of measurements we have also performed more complex operations
by putting OAM in both writing beams. The results are shown in Fig. 4 for a subset
of values of the  topological charge, and clearly demonstrates that one can store two OAM values using the CPO-based memory and then retrieve the sum of them. So, for example, starting from OAMs values  $\ell_{W}$=-2  and  $\ell_{W'}$=2 we have generated $\ell_{out}$=4. The diagonal of Fig. 4 illustrate the case of $\ell_{W}=\ell_{W'}$, retrieving a FWM output $\ell_{out}$=0. The non perfect reconstruction of the topological charge, specifically  for the $\ell_{out}=0$ case, come from the zero field value at the LG center inducing no FWM at this region. 
These results, however, are consistent with the conservation law of OAM within the field modes in this delayed
FWM process where photons are absorbed from the W and R modes and emitted into the
W' and D modes.

  \vspace{0.4cm}
In conclusion, we have demonstrated the storage and retrieval of OAM of light
through a CPO based memory in cold cesium atoms. A time delayed FWM configuration
was employed and the retrieved OAM was shown to be governed by the conservation
law of OAM into the incident field modes. We have demonstrated that this memory can
be used to perform logical operations involving the stored information encoded in
OAM. Our results can be used, for example, to implement a CNOT gate, \cite{Li14},
with memory, using the CPO mechanism. Finally, one may think to extend the CPO memory to the single-photon level, where the single-photon information is transferred to the external degree of freedom of an atom (momentum) and indistinguishably distributed among all the atoms in the ensemble.  
   
We acknowledge D. Bloch for stimulating discussions.
This work was supported by the Brazilian agencies CNPq and FACEPE. We also thank CAPES-COFECUB (Ph 740-12) for the support of Brazil-France cooperation.\indent

\emph{Permanent addresses}

\emph{$(\dagger)$ Instituto de Fisica, Facultad de Ingenier'a, Universidad de la Republica, J. Herrera y Reissig 565, 11300 Montevideo, Uruguay}

\emph{($\ddagger$) Unidade Academica do Cabo de Santo Agostinho, Universidade Federal Rural
de Pernambuco, BR 101 SUL, Km 97 - S/N, Cabo de Santo Agostinho,
PE, Brazil.}

\pagebreak

\section*{Informational Fifth Page}

\end{document}